# Holo-Television System with a Single Plane

José J. Lunazzi*[1], Daniel S. F. Magalhães[1], Noemí I. R. Rivera[1], Rolando L. Serra [2]

1 Laboratório de Óptica, Instituto de Física Gleb Wataghin, P.O.Box 6165, University of Campinas – UNICAMP, 13083-970 Campinas, SP, Brazil
2 Departamento de Física, Instituto Superior Politécnico "José Antonio Echeverría" Ave. 114, 11901, Marianao, Ciudad de La Habana, CP 19390, Cuba
*Corresponding author: lunazzi@ifi.unicamp.br

## Abstract

We show a system capable of projecting a video scene on a white-light holographic screen to obtain a kind of image that results in a plane in front of the screen. This holographic screen is mainly a diffractive lens and it is constructed by holography. The image plane can be located at any azimuth angle and seen with continuous parallax and without the use of goggles or any special visualization equipment. The image is not volumetric but when the plane is oblique to the observer its appearance looks very close to a real volumetric image.

An effort to develop a three-dimensional television system has been made in the last few years. 3D TV is being broadcasted in Japan for the first time in the world. To be viewed the observer must have a special LCD receptor and wear specific goggles, as with the Hyundai E465S 3D Stereoscopic LCD TV. Also, the regular use of these goggles has been shown to cause fatigue [1]. Goggle-less 3D monitors did not have any TV capability yet. A European project on 3DTV capture has been functional since September 2004 [2] but does not point to any display system. In Korea [3] a constant effort concerning online gaming and 3D animation is being gradually increased. 3D systems constructing an interference fringe pattern structure to be diffracted do not report representing a TV scene [4-6]. Alternative systems proposed present the inconvenience of needing to track the observer's position [7], so being valid for a single observer. Displays made of photorefractive polymers [8,9] offer a dynamic holographic recording material that allow updating of images but at very low frame rate wherein each complete static scene needs 2.5 minutes to be recorded, making their applications impossible with an animated scene.

Holographic screens had its applications extended to multiple projection and white-light systems [10,11]. This kind of holographic screen allows the decoding on white-light of a previous encoded image [12-14]. A previous system described in [15] shows the capability to transmit TV images to be viewed in continuous horizontal parallax only (HPO) without needing any goggles or information on the observer's position, for up to three simultaneous observers at different vertical levels. The HPO is an effective approximation to full-parallax imaging, because humans perceive depth using horizontally offset eyes. Also a volumetric computer image was created by slicing a 3D animation into planes that appear traversing the screen [15]. The planes were sequentially laterally displaced by means of a rotating mirror mounted on a high-speed computer controlled step motor. Extracting slices of 3D animated images from computer-generated images tested this system. When reducing the many planes to only one, the possibility of using a simplified TV system which projects a single plane was evident to us. The image did not look like a strictly plane image. Because depth is present, the visual sensation helps the observer to recognize the elements of a volumetric image.

In this Letter we describe a simplified Holo-Television system based on a first diffraction processes at a grating and a second one at a holographic screen. The holographic screen is characterized, the depth of the image is measured and some important results for this kind of holo-television system are presented.

The holographic screen is a high-resolution diffractive screen constructed by means of holographic recording; it is mainly an off axis diffractive lens in which the lens action is employed to define very directionally the viewing position, thus defining a multiple scene capability, each scene corresponding to a different viewing position. The observer's field is extended by the action of a thin diffuser, which is the object holographically registered [11]. The scheme to record a holographic screen is showed in Figure 1. The laser (L) emits a beam that is divided by a beam splitter (BS). One of the beams is reflected by a plane mirror (M) and reaches a spatial filter (SP) causing the beam to diverge. The other beam reaches a cylindrical lens (CL) of focal length 3.8 mm and it makes the beam to diverge creating a diffusion line at the diffuser (D). The angle between the two involved beams was 45 degrees. Each point of the diffuser illumines the entire holographic film (HF), where the interference pattern is recorded. The holographic film employed was a holographic silver halide film Agfa 8E75 HD. When the holographic screen (HS) is illuminated by a diverging white-light beam, each wavelength will converge at a different lateral position and distance because it makes the diffuser's image, as shown in Figure 2. The diffuser being only 2 mm wide, if the right eye of the observer is located precisely at the position of the green image of the diffuser, receiving light at the same wavelength from the whole screen, his left eye will receive light corresponding to a wavelength displaced towards the blue which may come from the whole screen area because it is composed of the many beams converging towards the eye. The purpose of the use of the diffuser at recording is that it vertically extends the position in which the observer can see the entire scene.

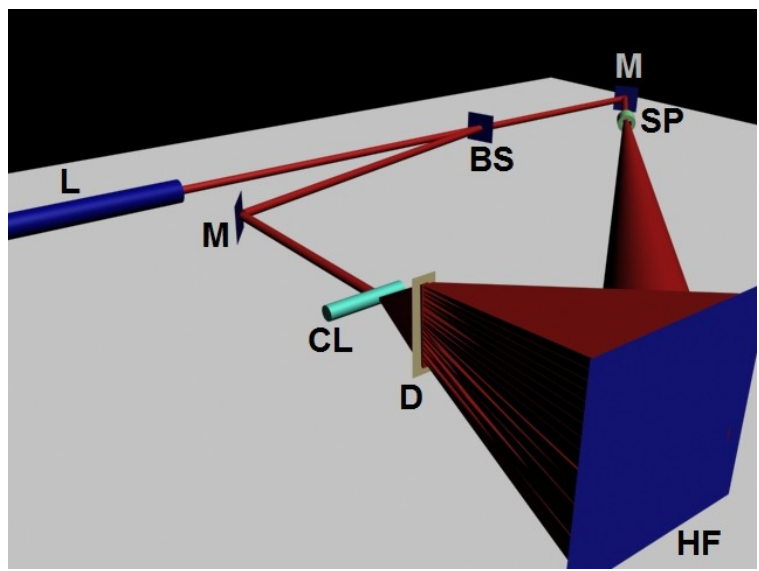

Fig. 1. The scheme to record a holographic screen. The laser (L) emits a beam which is divided by a beam splitter (BS). One of the beams is reflected by a plane mirror (M) and reaches a spatial filter (SP) causing the beam to diverge. The other beam reaches a cylindrical lens (CL) of focal length 3.8 mm and it makes the beam to diverge creating a diffusion line at the diffuser (D). The interference of both beams is recorded at the holographic film (HF).

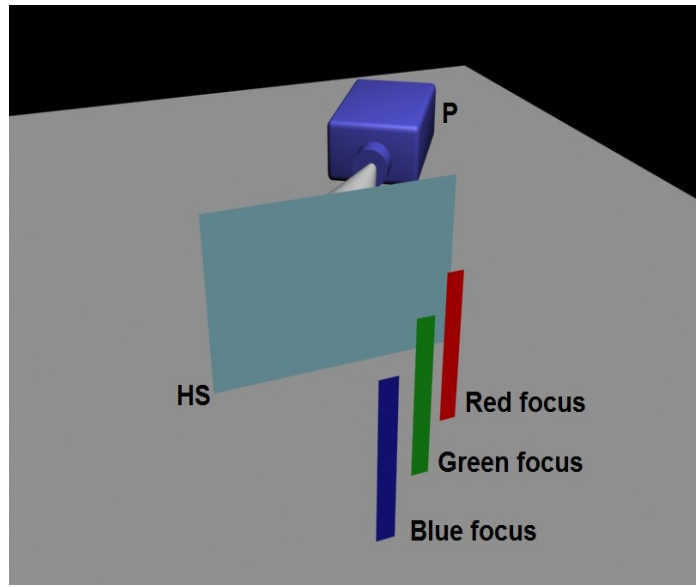

Fig. 2. Holographic screen being illuminated by a diverging white-light beam.

Using a diffraction grating as a spectral encoder of depths to project a 2D image, this image can be located in front, behind [13] or oblique [16] to the screen. Figure 3 shows the schematic settings to generate a floating image. A personal computer with resolution set to 800x600 pixels sends the image to an ordinary 2D multimedia projector (PJ) and projects the image on the diffraction grating. The projector had its lens inverted to reduce aberrations when focusing the image closer. The image is focused close to the diffraction grating (DG). The first diffraction order is caught by an objective lens (OL) with great aperture and projected on the holographic screen. This diffraction order appears like a spectral blur on the screen. The width of this spectral blur is the consequence of the distance between focused image and diffraction grating and it determines the depth in which the observed image will appear. The holographic screen diffraction makes the imaging process complete [14].

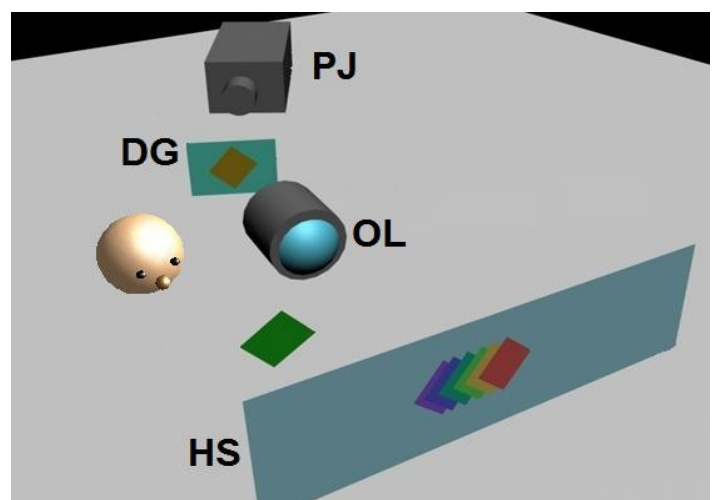

Fig. 3. 2D projection of an image in front of the holographic screen (HS). A multimedia projector (PJ) projects a 800x600 pixels image on a diffraction grating (DG) of 1000 lines/mm and an objective (OL) 1:2.8 with 90 mm of focal length focalizes the first diffraction order of DG on HS. The distance image-HS is (27.0±0.5) cm.

The system employs a Sharp XG-H400U multimedia projector, a diffraction grating of 1000 lines/mm with diffraction efficiency (13±1)% and a Pentax Takumar objective 1:2.8 with 90 mm of focal length. The holographic screen used has a focal length of (64±1)cm, a diffraction efficiency of (23±1)% and 30x60 cm of size. The involved distances are PJ-DG=21.5 cm, DG-OL=8 cm and OL-HS=171.8 cm.

The image can be observed in two different ways: the reflected diffraction order and the transmitted diffraction order of the holographic screen. The diffraction efficiency in both cases is almost the same. In our specific case we choose the reflected order to optimize the available room space. Figure 4 shows a frame of a recorded video with a digital video camera and two pictures of the image projected with the system [18]. The center and right pictures of Fig. 4, were captured using a non-professional digital camera Olympus X-760, the photographic result being poorer than the effect experienced upon direct viewing.

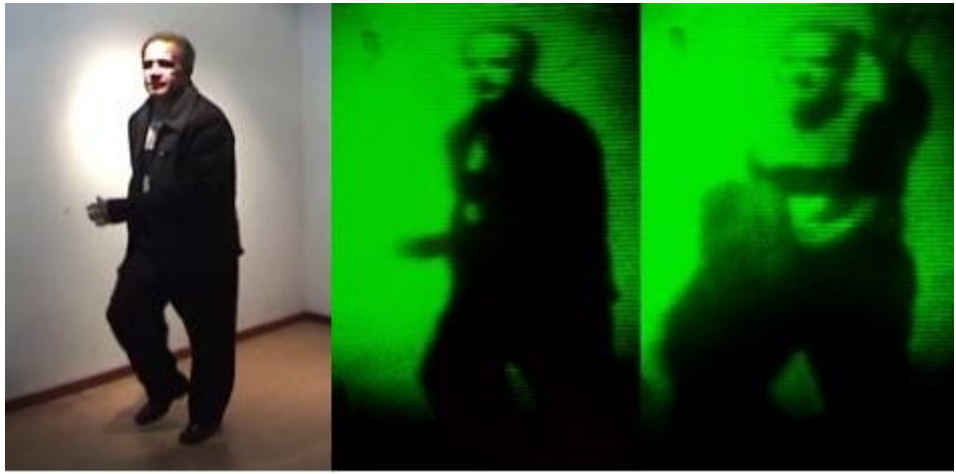

Fig. 4. At left the visualization of the projected video scene. At center the visualization of this frame projected on the holographic screen. At right another part of the same projected video. The video can be seen at [18]

To find the position of the image we use a stick as a reference. The camera is put at the position of each eye of the observer. Displacing the stick, when it is at the same line that a selected image point for both viewpoints, the position of the image is found.

Figure 5 shows the position when both eyes observe the stick aligned with the same image of the legs. The corresponding z distance between the holographic screen and the image of the man was (27.0±0.5)cm in front of the screen. The observer's field of view at 140 cm from the screen has (24±1)cm which corresponds to an angular field of 11±1°

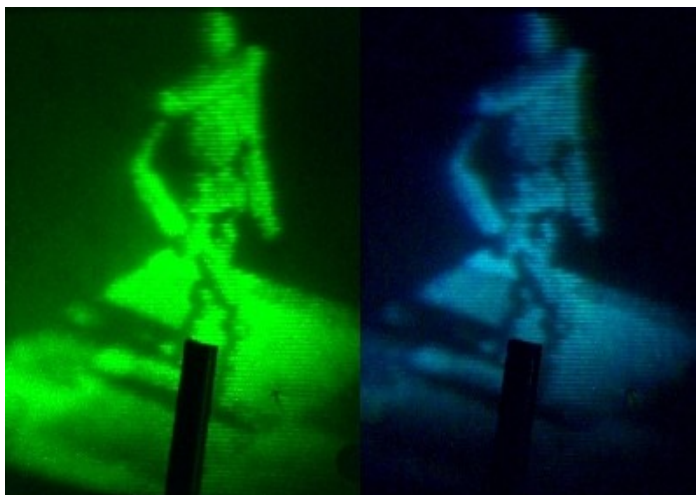

Fig. 5. At left, the left eye vision of the image and at right, the right eye vision of the image. The comparison between the stick and the image shows that the image is 27 cm in front of the screen.

The observed image has continuous horizontal parallax and because the wavelength change is not enough to keep the same color sensation, in each observation point the image appears in a different color, the left image of Fig. 5 can be seen in green and the right image can be seen in blue [18,19].

In summary, a holographic screen working on white light was characterized by means of the knowledge of its spatial frequency. We present a system capable of projecting a computer animation or a video scene on a white-light holographic screen to obtain an image with continuous horizontal parallax that results in front of the screen. Color images could be obtained using three projectors in RGB synchrony overlapped at the same holographic screen [17].